\documentclass[twocolumn,floatfix,aps,prd,noshowpacs,footinbib,amsmath,amssymb,amsfonts]{revtex4}

\usepackage{graphicx}

\usepackage{bm}
\newcommand{\be}{\begin{equation}}
\newcommand{\ee}{\end{equation}}
\newcommand{\ba}{\begin{eqnarray}}
\newcommand{\ea}{\end{eqnarray}}

\begin{document}

\title{Disentangling non-Gaussianity, bias and GR  effects in the galaxy distribution}

\author{Marco Bruni$^1$, Robert Crittenden$^1$, Kazuya Koyama$^1$,
Roy Maartens$^{2,1}$, Cyril Pitrou$^1$, David Wands$^1$}

\affiliation{$^1\,$Institute of Cosmology \& Gravitation,
University of Portsmouth, Portsmouth PO1 3FX, United Kingdom \\
$^2\,$Department of Physics, University of Western Cape, Cape Town
7535, South Africa}

\begin{abstract}
Local non-Gaussianity, parametrized by $f_{\rm NL}$,  introduces a
scale-dependent bias that is strongest at large scales, precisely where General Relativistic (GR) effects also become significant. With future data, it should be possible to constrain
$f_{\rm NL} = {\cal O}(1)$ with high redshift surveys.
GR corrections to the power spectrum and ambiguities in the gauge used to define bias introduce effects similar to $f_{\rm NL}= {\cal O}(1)$, so it is essential to disentangle these effects.
%*
For the first time in studies of primordial non-Gaussianity, we include the consistent GR calculation of galaxy power spectra, highlighting the importance of a proper definition of bias.
We present observable power spectra with and without GR corrections, showing that an incorrect definition of bias can mimic non-Gaussianity. However, these effects can be distinguished by their different redshift and scale dependence, so as to extract the true primordial non-Gaussianity.
%*
%We show how to consistently include
%primordial non-Gaussianity in the observed angular power spectrum of %the galaxy distribution and we discuss how to distinguish between %the various effects, so as to extract an accurate non-Gaussianity %signal.

\end{abstract}

\maketitle

\section{Introduction}
The galaxy power spectrum has proven an indispensable tool in understanding our Universe, and measuring it underlies the motivation for many future surveys.  Usually the 3D power spectrum $P(k)$ is inferred from the measured galaxy redshifts and angular positions and compared to predictions based on constant-time slices of the Universe.
However, surveys are slices down the past lightcone of the observer; constant-time slices are in principle unobservable for significant redshifts and, for a given observer, are not uniquely associated with redshift.
Furthermore, for large-scale perturbations  of the order of the Hubble radius $H^{-1}$,  $P(k)$ needs to be defined for the matter overdensity $\delta_m$ in a given choice of time slice, i.e.\ a choice of coordinates or ``gauge", which can be arbitrarily specified. On sub-Hubble scales, the power spectrum is not sensitive to the choice of gauge, but significant ambiguities arise near and beyond  $H^{-1}$. It is possible to define a gauge-invariant $\delta_m$, but this
does {\em not} solve the problem.
Instead, one must compare predictions for true observables, such as angular power spectra at a given redshift, which are necessarily gauge invariant.

In order to relate the observed overdensity to the overdensity defined in any chosen gauge, General Relativistic (GR) corrections must be included to take account of both gauge effects and lightcone effects. These corrections have recently been analyzed in detail \cite{Yoo:2009au,Yoo:2010ni,Bartolo:2010ec, Jeong,Bonvin,Challinor}. The corrections are relevant close to and beyond the Hubble scale $H^{-1}$. For higher-redshift surveys, the corrections also become significant on smaller scales, since $H^{-1}(z)$ decreases with the redshift $z$.
For example,
$H^{-1}(z=2) \sim 0.4 H_0^{-1}$, and $H^{-1}(z=10) \sim 0.05 H_0^{-1}$.  %
The angular extent of the Hubble radius at redshift $z$ is $\theta_H(z)\equiv {H^{-1}(z)}/{D_A(z)}$, where $D_A(z)$ is the angular diameter distance.
At $z = 2$, we have $\theta_H \sim 50^\circ$.

Just as the overdensity definitions depend on the gauge, so too does the galaxy bias, which relates the fluctuations of galaxy number density to the underlying matter density fluctuation.
An ambiguity in the definition of bias can not only confuse our understanding of the underlying density power spectrum, but it also can impact our understanding of the non-Gaussianity of the underlying density field, because local type non-Gaussianities characterised by $f_{\rm NL}$ have been shown to produce a scale-dependent bias \cite{Dalal:2007cu, Slosar:2008hx} (for a review see Ref.~\cite{Desjacques:2010jw} and references therein).

On large scales, measurements are limited by cosmic variance.
However,
a recent proposal aims to eliminate cosmic variance by correlating a highly biased tracer of large-scale structure against an unbiased tracer
\cite{Seljak:2008xr}. Using this method,
constraints at the level
$f_{\rm NL} = {\cal O}(1)$ are estimated \cite{Hamaus:2011dq}. The GR corrections and the gauge dependence of the bias
introduce effects in the observed angular power spectra of the same
order as the scale dependent bias effects with $f_{\rm NL} = {\cal O}(1)$,
as we show below.
Thus it is essential to include the GR corrections and clarify the gauge dependence of the bias prescriptions in order to obtain optimal constraints on primordial non-Gaussianity.

\section{Prescriptions for galaxy bias}
It is generally
assumed that the galaxy bias is scale-independent on large scales if the primordial fluctuations are Gaussian. However, once  GR corrections become important, it is essential to specify the physical frame in which the scale-independent bias is defined.
In \cite{Yoo:2009au,Yoo:2010ni,Bartolo:2010ec} the scale-independent bias is defined in the uniform redshift gauge, i.e.\ it relates the observed galaxy overdensity to the matter density in the same observational gauge. However, in general  bias is
determined by local physics \cite{Coles} that governs the formation of galaxies and
is independent of how they are observed.

Local bias should be defined in
the rest frame of CDM
(which is typically assumed to coincide with the baryon rest frame).
Bias can be understood in the peak-background-split approach where halo collapse occurs due to small scale peaks in the density exceeding a critical value, $\delta_\star$ \cite{Kaiser:1984sw}. Large-scale fluctuations $\delta_{\rm long}$ lead to an enhancement of the density of collapsed halos by effectively reducing the critical value for small-scale fluctuations to $\delta_\star-\delta_{\rm long}$.
In the spherical halo collapse model, there is an exact GR interpretation of the standard Newtonian collapse  \cite{Wands:2009ex}.
The criterion for collapse of a local overdensity is that the linear overdensity in the comoving-synchronous gauge reaches the critical value, $\delta_\star=1.686$.
Thus linear scale-independent
bias applies on large scales for a Gaussian distribution in the comoving-synchronous gauge
(as used by \cite{Jeong,Bonvin,Challinor}).
By contrast, in the Newtonian (longitudinal) gauge there is no local criterion for collapse in terms of the
overdensity on large scales \cite{Wands:2009ex}.

In comoving-synchronous gauge
(ignoring vector and tensor modes)
\be
ds^2 = a(\eta)^2 \Big\{-d\eta^2 + \Big[ (1-2 {\cal R}) \delta_{ij} + 2 \partial_i \partial_j E \Big] dx^i dx^j \Big\},
\ee
and the CDM velocity perturbation vanishes in this gauge.
In the simple linear model, the Eulerian
bias $b$ is defined via
\begin{equation}
 \label{defb}
 \delta_n^{(c)} = b \delta_m^{(c)} \,,
\end{equation}
where $\delta_n^{(c)}=\delta n^{(c)}/n$ and $\delta_m^{(c)}=\delta\rho^{(c)}/\rho$  are  the galaxy and matter fractional  overdensities in comoving-synchronous gauge.

Although the observed galaxy overdensity can be calculated in any gauge, it  is often convenient to use the Newtonian gauge because it facilitates comparisons with the standard Newtonian calculations.
In Newtonian gauge,
\begin{equation}
ds^2= a(\eta)^2 \Big[- (1+2 \Phi) d \eta^2 + (1-2\Psi) d \bm{x}^2 \Big].
\end{equation}
In the matter-dominated era, we can set $\Psi=\Phi$ since the effect of radiation anisotropic stress is negligible. From now on, quantities without superscripts refer to Newtonian gauge values if there is ambiguity. The density perturbations in
the two gauges are related by
\begin{equation}
 \delta_n^{(c)} = \delta_n - \frac{n'}{n}\frac{v}{k} \,, \quad \delta_m^{(c)} = \delta_m -\frac{\rho'}{\rho}\frac{v}{k}  \,,
 \label{gauge}
\end{equation}
where $v$ is the velocity.
Thus in Newtonian gauge, using
(\ref{defb}) we have \mbox{$\delta_n = b \delta_m - \left[ b (\ln \rho)' - (\ln n)' \right] v/k$}.
Assuming that galaxy number and matter are both conserved, we have $n'/n=\rho'/\rho=-3 a H$ and hence
\begin{equation}
 \delta_n = b \delta_m^{(c)} - 3\frac{a H}{k }  {v} \, = b \delta_m + 3\frac{a H}{k } (b-1) {v} \,.
 \label{gal-Newton}
\end{equation}
On small scales, $a/k \ll H^{-1}$, the gauge difference vanishes and $\delta\simeq \delta^{(c)} \simeq \delta^{{\rm std}}$, where $\delta^{{\rm std}}$ is the standard density perturbation in the Newtonian limit.
Note that for the growing mode in the matter era (i.e.\ neglecting any dark energy effect), in Newtonian gauge we have
\begin{equation}
\delta_m = -2 \left[ 1 + \frac{k^2}{3{(a H)}^2} \right] \Phi \,, \quad 3{a H} \frac{v}{k} = 2\Phi \,,
\end{equation}
where the Newtonian potential $\Phi$ remains constant on all scales.
In comoving-synchronous gauge
\begin{equation}
\delta_m^{(c)} = - \frac{{2k^2}\Phi}{{3{(a H)}^2}}.
\label{poisson}
\end{equation}
The two gauges coincide on sub-Hubble scales, but they differ significantly near the Hubble scale.

In \cite{Jeong,Bonvin,Challinor}, the analysis
of GR corrections to the power spectrum
is performed for a Gaussian primordial spectrum. Here we generalize to include primordial non-Gaussianities of the local form. The potential $\Phi$
has a non-Gaussian contribution:
\begin{equation}
\Phi = \phi_g - f_{\rm NL} \big(\phi_g^2 - \langle \phi_g^2 \rangle \big),
\label{local}
\end{equation}
where $\phi_g$ is Gaussian. The Newtonian potential $\Phi$ is related to the density perturbation in the comoving-synchronous gauge by the Poisson equation (\ref{poisson}).
Therefore this local form for the Newtonian potential corresponds to a non-local form
for the density field in the comoving-synchronous gauge.
Following \cite{Dalal:2007cu, Slosar:2008hx}, this non-Gaussianity induces a scale-dependent correction to the bias:
\be
b=\bar{b}+f_{\rm NL}(\bar{b}-1) {3\delta_\star \Omega_m H_0^2 \over k^2 T(k) D(a)},
\label{non-G-bias}
\ee
where $\bar b$ is the Gaussian bias,
$T(k)$ is the transfer function ($k$-independent at low $k$),  and $D(a)$ is the growing mode of density perturbations. We emphasize that this definition is intrinsically relating the comoving-synchronous gauge densities \cite{Wands:2009ex}.

In this paper, we only include local-type primordial non-Gaussianity (\ref{local}) for simplicity. However, any
non-Gaussianity that produces a non-zero bispectrum of the Newton potential
in the squeezed limit in the matter era leads to a scale dependent correction
to the bias \cite{Matarrese:2008nc}.
This includes non-linear evolutions of the potential on super-horizon scales
\cite{Verde:2009hy} and it should be added to (\ref{non-G-bias}).

%%%%%%%%%%%%%%%%%%%%%%%%%%%%%%%%%%%%%%%%%%%%%%%
\begin{figure*}[t]
\includegraphics[width=18cm]{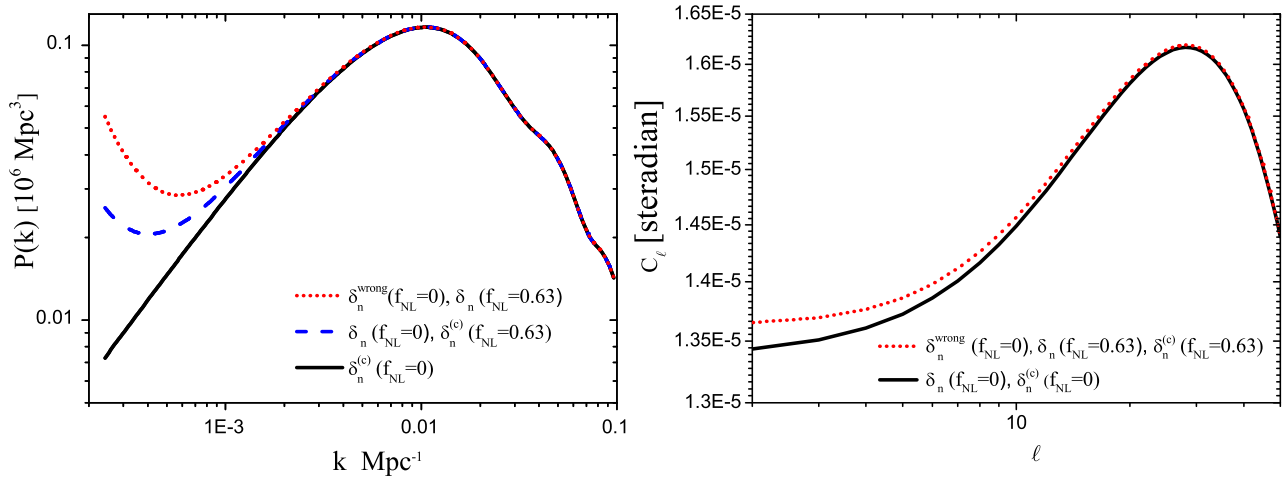}
\vspace{-5.8cm}
\caption{{\it Left:} The power spectrum of various density perturbations at $z=1$. We use a standard LCDM background and assume $\bar{b}=2$. {\it Right}: The angular power spectrum at $z=1$ assuming all galaxies are in a Gaussian window function of width $\sigma_z=0.1$.
See text for detailed explanation.
Note that we take $f_{\rm NL}=0.63$, which is slightly different from $f_{\rm NL}=3/A$ due to the effect of the cosmological constant. Also $A$ evolves with redshift and a different $f_{\rm NL}$ would be required at another redshift. }
\label{powspec}
\end{figure*}
%%%%%%%%%%%%%%%%%%%%%%%%%%%%%%%%%%%%%%%%%%%%%%%%
\section{Gauge dependence of the power spectrum}
Figure \ref{powspec} (left panel) illustrates how the number density power spectrum $P(k)$ depends on the gauge choice of the variables, on the gauge used to define the bias and on the non-Gaussianity. The simple linear behaviour (solid line) results from plotting the comoving-synchronous gauge number density using a linear scale-independent bias defined in comoving-synchronous gauge, (\ref{defb}).
The corresponding Newtonian gauge number density (dashed line) differs by a term ${\cal{O}}(k^{-2})$. This scale dependence is exactly the same as the scale dependence of the bias in the comoving-synchronous gauge arising from primordial non-Gaussianity (\ref{non-G-bias}).

Consider the galaxy overdensity $\delta_n$ given by (\ref{gal-Newton}) in the matter era. On large scales, the velocity term $3 aH v/k$ and the non-Gaussian correction to the bias (\ref{non-G-bias}) dominate and $\delta_n$ behaves as
\be
\delta_n \to \Big[3 + f_{\rm NL} A (\bar{b}-1) \Big] \frac{{(a H)}^2}{k^2} \delta_m^{(c)},
\label{gal-Newton-large}
\ee
where we defined
\be
A(z) = \frac{3\delta_\star \Omega_m}{T(k) D(z) a^2} \frac{H_0^2}{H^2(z)} =2
\delta_\star \frac{k^2}{(a H)^2}\frac{\Phi_{\rm init}}{\delta^{(c)}_m},
\ee
where $\Phi_{\rm init}$ is the potential in the radiation era.
The gauge correction to the galaxy overdensity in Newtonian gauge, corresponding to the first term in square brackets in (\ref{gal-Newton-large}), gives a similar ${\cal{O}}(k^{-2})$  effect as the non-Gaussian correction to the bias, the second term.
In the matter-dominated era, the power spectrum of the Newtonian gauge $\delta_n$ coincides with that in the comoving-synchronous gauge $ \delta^{(c)}_n$, with
$f_{\rm NL} = {3 / [A (\bar{b}-1)]}.$
We emphasize that this is an apparent effect -- if we calculate true observables, they are independent of the choice of gauge and there is no confusion. However, this clearly demonstrates the importance of using true observables that include all the GR corrections when analyzing the effect of primordial non-Gaussianity.

In addition to the gauge dependence of the galaxy
overdensity, the ambiguities in the gauge used to define bias introduce another potential confusion to the scale-dependent bias effect from primordial non-Gaussianity. The dotted line shows the impact on the Newtonian gauge
overdensity of wrongly defining a linear bias in Newtonian gauge, $\delta_n^{\rm wrong}=\bar{b} \delta_m$. On large scales, this behaves as
\be
\delta_n^{\rm wrong}  \to  3 \bar{b} \frac{{(a H)}^2}{k^2} \delta_m^{(c)}.
\label{bias-Newton}
\ee
Comparing (\ref{gal-Newton-large}) and (\ref{bias-Newton}), we find that if the linear bias is assumed to be scale-independent in  Newtonian gauge, this would be interpreted as the effect of primordial non-Gaussianity with $f_{\rm NL}=3/A$
in the matter-dominated era.

\section{Observed angular power spectra}
Figure~\ref{powspec} (left panel) illustrates how non-Gaussianity could be confused in the power spectrum with GR gauge corrections on large scales.
However, the power spectrum and the overdensities themselves are {\em not} directly observable. In the Newtonian limit, it is standard to define the observed galaxy overdensity as
\begin{equation}
\Delta^{{\rm std}}=
\delta^{{\rm std}}_n - \frac{1}{aH} \hat{\bm{n}} \cdot \frac{\partial \bm{v}}{\partial \chi}  -2 \kappa,
\label{std}
\end{equation}
where $\delta^{{\rm std}}_n$ is the number density fluctuations in the Newtonian limit,
$\bm{v}$ the peculiar velocity, and
$\kappa$ is the lensing convergence \cite{Challinor}
 \be
\kappa=-{1\over2}\nabla_{\hat{\bm{n}}}^2\int_{\eta_{\rm o}}^\eta {(\tilde \eta-\eta) \over (\eta_{\rm o}-\eta)(\eta_{\rm o}- \tilde\eta)}(\Phi+\Psi) d\tilde\eta.
 \ee
However, there is an ambiguity in the definition of $\delta^{{\rm std}}_n$ in this formula on large scales because we need to specify the gauge for $\delta^{{\rm std}}_n$. We must also include all effects impacting the observed galaxy overdensity, including redshift-space distortions, gravitational lensing and integrated Sachs-Wolfe effects.

The observed galaxy overdensity including these effects has been first derived in \cite{Yoo:2009au,Yoo:2010ni}. In these papers, bias is defined on the constant observed redshift hypersurface. As discussed in section III, bias should be defined in the comoving-synchronous gauge and a wrong definition of bias introduces spurious scale dependent effects in the power spectrum.
Refs.~\cite{Challinor, Bonvin} derived the observed galaxy overdensity in Newtonian gauge by correctly defining bias in the comoving-synchronous gauge. The observed galaxy overdensity in the direction $\hat{\bm{n}}$ at redshift $z$ is obtained as
\ba
&& \Delta_n(\hat{\bm{n}},z) = \delta_n - \frac{1}{aH} \hat{\bm{n}} \cdot \frac{\partial \bm{v}}{\partial \chi}  -2 \kappa \nonumber\\
&&~~+ \left(1+ {\dot H \over H^2}+\frac{2}{aH\chi}
\right)
\left[  \Phi + \int^{\eta_{\rm o}} ({\Phi}' + {\Psi}') d \eta - \hat{\bm{n}}\cdot \bm{v} \right] \nonumber
\\
&&~~+ \frac{2}{\chi}\int^{\eta_{\rm o}}(\Phi + \Psi)d\eta
+ \frac{1}{aH}{\Psi}' +\Phi- 2\Psi, \label {deltan}
 \ea
where all perturbations are expressed in Newtonian gauge,
a subscript `o' denotes the observer,  and $\chi$ is the comoving radial position of the source. The first three terms are the same as the standard formula in the Newtonian limit.
The second line comes from the redshift perturbation of the volume, which contains a Doppler term, the ordinary and integrated Sachs-Wolfe terms. The integral in the third line comes from the radial shift due to lensing. The potentials, $\Psi$ and $\Phi$, are suppressed on small scales compared with density fluctuations but they become the same order near the horizon scales and their effects need to be taken into account consistently.

%%%%%%%%%%%%%%%%%%%%%%%%%%%%%%%%%%%%%%%%%%%%%%%%
\begin{figure*}[t]
\includegraphics[width=18cm]{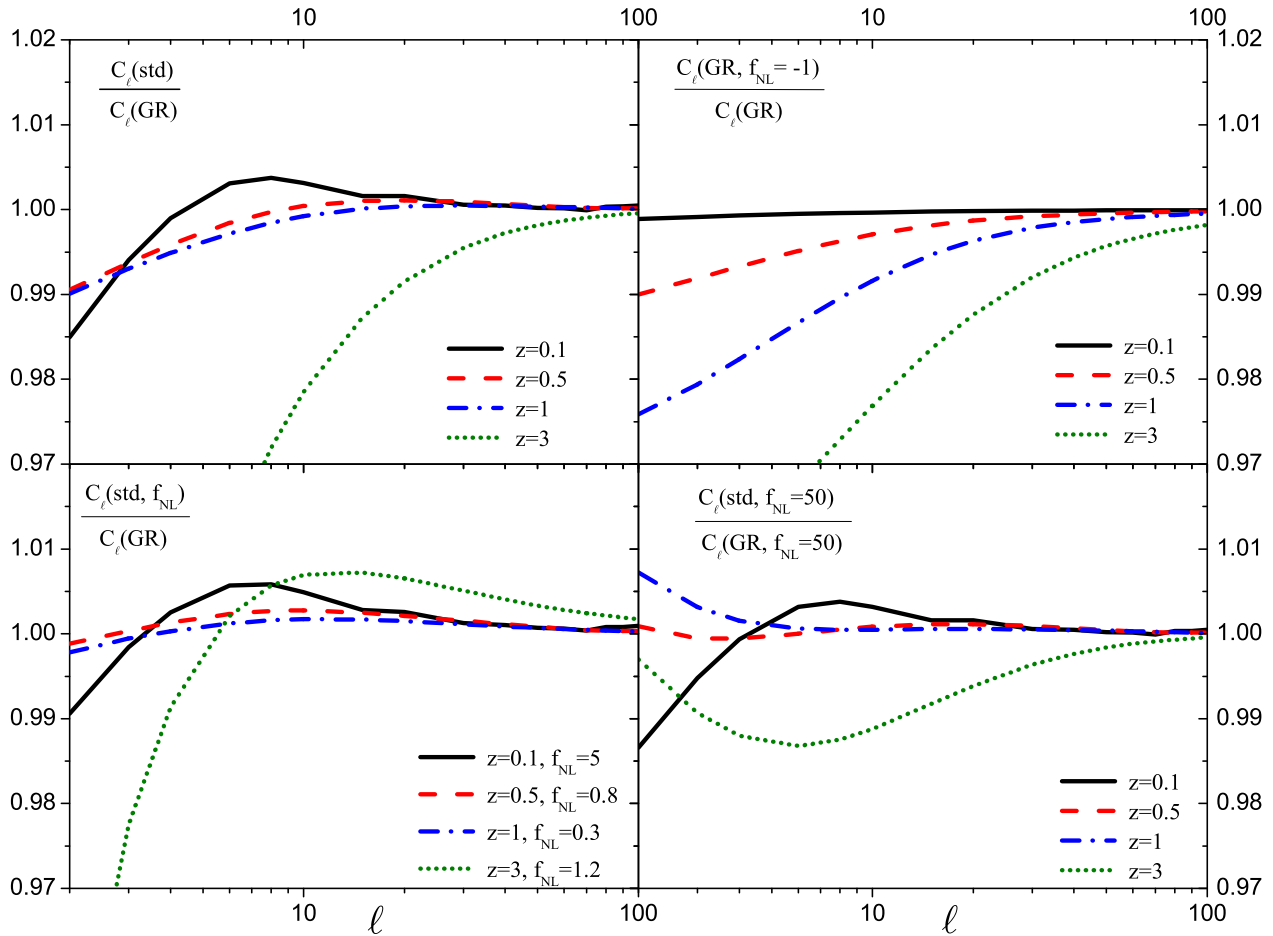}
\caption{Ratio of the standard angular power spectra to the full GR spectrum in various cases. See text for detailed explanation.
We assume $\bar{b}=2$ and all the galaxies are in a Gaussian window function of width $\sigma_z=0.1 z$. \label{powspec2}}
\end{figure*}
%%%%%%%%%%%%%%%%%%%%%%%%%%%%%%%%%%%%%%%%%%%%%%%%
In order to compute the observable effects of GR corrections and non-Gaussianity, we consider the angular power spectrum $C_\ell$ of $\Delta_n$ in a redshift slice.
The counts in a redshift slice are given by
 \be
\Delta_n(\hat{\bm{n}},z)=\sum_{\ell,m}a_{\ell m}(z) Y_{\ell m}(\hat{\bm{n}}), ~~ C_\ell(z)= \langle |a_{\ell m}(z)|^2 \rangle.
 \ee
This angular power spectrum includes the auto- and cross-correlations amongst all terms on the right of (\ref{deltan}), taking into account the relevant transfer functions \cite{Bonvin}.

Figure~\ref{powspec} (right panel) shows the angular power spectrum at $z=1$ assuming all galaxies are in a Gaussian window function of width $\sigma_z=0.1$. As emphasized before, the observed angular power spectrum is independent of gauge, so that there is no ambiguity in the choice of gauge for the galaxy density perturbation. However, there still remain the ambiguities in the gauge used to define bias. The dotted line shows the result of defining a linear bias in the Newtonian gauge, which is degenerate at fixed redshift with a scale-dependent bias effect from primordial non-Gaussianity.

In Figure~\ref{powspec2},
we compare the full angular spectra with the standard formula based on the Newtonian limit results, Eq.~(\ref{std}). As mentioned above, there is an ambiguity in the definition of $\delta^{{\rm std}}_n$ in this formula on large scales. We use the definition $\delta^{{\rm std}}_n = b \delta^{(c)}_m$ because the growth rate of $\delta^{(c)}_m$ is the same as the Newtonian result on all scales and the gauge difference is suppressed by the $1/k$ factor in Eq.~(\ref{gal-Newton})

 \cite{Challinor}.

The top-left plot shows the impact of full GR corrections on observed
spectra compared to the standard result at various redshifts, for
Gaussian fluctuations. At higher redshift, the GR effect becomes larger
and more important at larger $\ell$, since the Hubble scale is smaller.
These corrections are suggestive of the changes which arise due to
primordial non-Gaussianity, which are shown in the top-right panel,
including all GR corrections. However, at high redshifts, we find
that the shape of the angular power spectra induced by the GR
corrections is very different from the shape induced by the non-Gaussian
effect, and these two effects look very different at small $\ell$.
This is demonstrated in the bottom-left panel, where we attempted to
tune $f_{\rm NL}$ in the standard formula so that it cancels GR
corrections.  This lack of degeneracy implies that it is in principle
possible to disentangle between the two effects in the observed angular
power spectrum at different redshifts. The bottom-right panel shows
the GR corrections in the presence of primordial non-Gaussianity $f_{\rm NL}=50$.
At low redshifts, the non-Gaussian effect is still small and GR corrections are of the
same order as the Gaussian case. As expected, GR corrections become
unimportant for small $\ell$ and at high redshifts where the
non-Gaussian effects dominate. (All numerical computations have been
performed with the freely available package {\tt CMBquick}
\cite{CMBquick}.)

Note that in our analysis we have considered the simplest case of conserved galaxy number and neglected selection effects, e.g., luminosity dependence in a realistic magnitude-limited survey \cite{Bonvin,Challinor}. It is also interesting to consider possible degeneracies with other physical effects that have an influence on the power spectrum on large scales, such as clustering dark energy, tilted spectrum and isocurvature
modes, which would need further investigation.

\section{Conclusions}
%*
We observe on our past lightcone and the correct interpretation of data on very large scales requires GR corrections to the standard Newtonian approach.
This is particularly important for detecting non-Gaussianity using large-scale structure.
For the first time, we have included all GR corrections to the galaxy power spectrum in an analysis of primordial local-type non-Gaussianity. We only considered local-type primordial non-Gaussianity (\ref{local}) in this paper because any other types of primordial non-Gaussianity that produce a negligible bispectrum of the Newton potential
in the squeezed limit do not lead to a scale dependent correction \cite{Matarrese:2008nc}.
We have shown that galaxy bias is naturally defined in the comoving-synchronous gauge, and choosing a different gauge can lead to incorrect conclusions on scales approaching the Hubble radius where GR corrections become important.

GR corrections to the power spectrum and ambiguities in the gauge used to define bias introduce effects qualitatively similar to local-type non-Gaussianity with $f_{\rm NL}= {\cal O}(1)$. However we have shown, using the observed angular power spectrum, that while the wrong bias definition (\ref{bias-Newton}) has the same scale-dependence as non-Gaussian effects, it has different redshift dependence.Moreover, GR effects have a somewhat different scale dependence. These differences are the basis for discriminating between these effects and identifying the true non-Gaussian signal.

GR effects are important at wide angles and at high redshifts as these effects are important near horizon scales (see Fig.~1 and Fig.~2). Thus a deep and wide angle survey makes the effects easier to see. On large scales, measurements are limited by cosmic variance. Given that GR corrections are comparable to local-type non-Gaussianity with $f_{\rm NL}= {\cal O}(1)$, they are not measurable without reducing cosmic variance. However, a recent proposal aims to eliminate cosmic variance by correlating a highly biased tracer of large-scale structure against an unbiased tracer \cite{Seljak:2008xr}. It is important to study whether general relativistic effects are measurable in future surveys \footnote{Recently detectability of the velocity term is discussed in J.~Yoo, N.~Hamaus, U.~Seljak and M.~Zaldarriaga, arXiv:1109.0998 [astro-ph.CO]. }.

\[ \]{\bf Acknowledgments:}
We are supported by the STFC (grant no. ST/H002774/1); KK is
supported by a European Research Council Starting Grant and the Leverhulme trust; RM is supported by a South African SKA Research Chair. We thank Daniele
Bertacca, Ruth Durrer, Sabino Matarrese, Nikolai Meures, Irene Milillo, Misao Sasaki and
Licia Verde for helpful discussions. KK thanks Donghui Jeong and Fabian Schmidt for
useful discussions at the Cosmological Non-Gaussianity Workshop, University of Michigan.

\end{document}